\documentclass[journal]{IEEEtran}
\IEEEoverridecommandlockouts
\usepackage{booktabs}
\usepackage{siunitx}
\usepackage{cite}
\usepackage{amsmath,amssymb,amsfonts}
\usepackage{graphics, framed}
\usepackage{graphicx, enumerate}
\usepackage{epsfig}
\usepackage{epstopdf}
\usepackage{url}
\usepackage{multimedia}
\usepackage{paralist}
\usepackage[printonlyused]{acronym}
\usepackage{units}
\usepackage{algorithmic}
\usepackage{textcomp}
\usepackage{float}
\usepackage{tabularx}
\usepackage{balance}
\usepackage{subfigure}
\usepackage{xcolor}
\def\BibTeX{{\rm B\kern-.05em{\sc i\kern-.025em b}\kern-.08em
T\kern-.1667em\lower.7ex\hbox{E}\kern-.125emX}}
\usepackage{mathtools}
\usepackage{cuted} 
\usepackage[export]{adjustbox}

\usepackage{dblfloatfix}
\usepackage{lipsum}
\usepackage{color,soul}
\usepackage{multirow}
\usepackage[overload]{textcase}
\newcommand{\FGR}[1]{Fig.~\ref{#1}}

\newcommand{\SEC}[1]{Section~\ref{#1}}
\newcommand{\TAB}[1]{Table~\ref{#1}}
\acrodef{5G}[5G]{5\textsuperscript{th}-Generation}
\acrodef{BW}[BW]{bandwidth}
\acrodef{BER}[BER]{bit error rate}
\acrodef{IP}[IP]{Internet protocol}
\acrodef{BPSK}[BPSK]{binary phase-shift keying}
\acrodef{CW}[CW]{continuous wave}
\acrodef{CSI}[CSI]{channel state information}
\acrodef{D2D}[D2D]{device-to-device}
\acrodef{dB}[dB]{decibel}
\acrodef{dBi}[dBi]{decibel isotropic}
\acrodef{dBm}[dBm]{decibel over a milliwatt}
\acrodef{DSN}[DSN]{deep-space networks}
\acrodef{DTN}[DTN]{delay tolerant network}
\acrodef{Gbps}[Gbps]{gigabit per second}
\acrodef{GHz}[GHz]{gigahertz}
\acrodef{GAT}[GAT]{graph attention network}
\acrodef{THz}[THz]{Terahertz}
\acrodef{ISL}[ISL]{inter-satellite link}
\acrodef{RIS}[RIS]{Reconfigurable intelligent surfaces}
\acrodef{GM}[GM]{Gamma mixture}
\acrodef{PSK}[PSK]{phase shift keying}
\acrodef{QAM}[QAM]{quadrature amplitude modulation}
\acrodef{AWGN}[AWGN]{additive white Gaussian noise}
\acrodef{SNR}[SNR]{signal-to-noise ratio}
\acrodef{AF}[AF]{amplitude-and-forward}
\acrodef{MIMO}[MIMO]{multiple-input multiple-output}
\acrodef{mMIMO}[mMIMO]{massive-multiple-input multiple-output}
\acrodef{SDN}[SDN]{Software-defined network}
\acrodef{SON}[SON]{self-organizing network}
\acrodef{HetNet}[HetNet]{heterogeneous network}
\acrodef{FSO}[FSO]{free-space optics}
\acrodef{UM-MIMO}[UM-MIMO]{ultra-massive-MIMO}
\acrodef{AP}[AP]{access point}
\acrodef{UE}[UE]{user equipment}
\acrodef{NTN}[NTN]{non-terrestrial networks}
\acrodef{UAV}[UAV]{unmanned aerial vehicle}
\acrodef{HAPS}[HAPS]{high-altitude platform station}
\acrodef{LEO}[LEO]{low-Earth orbit}
\acrodef{BAN}[BAN]{body area network}
\acrodef{WLAN}[WLAN]{wireless local area network}
\acrodef{QoS}[QoS]{quality of service}
\acrodef{TCS}[TCS]{thermal control system}
\acrodef{QCL}[QCL]{quantum cascade laser}
\acrodef{CMOS}[CMOS]{complementary metal-oxide semiconductor}
\acrodef{V-HetNet}[V-HetNet]{vertical heterogeneous network}
\acrodef{DL}[DL]{deep learning}
\acrodef{DRL}[DRL]{deep reinforcement learning}
\acrodef{EIRP}[EIRP]{effective isotropic radiated power}
\acrodef{FDTD}[FDTD]{Finite-difference time-domain}
\acrodef{FEM}[FEM]{finite element method}
\acrodef{MoM}[MoM]{method of moments}
\acrodef{VNA}[VNA]{vector network analyzer}
\acrodef{CS}[CS]{channel sounder}
\acrodef{CIR}[CIR]{channel impulse response}
\acrodef{CTF}[CTF]{channel transfer function}
\acrodef{DPM}[DPM]{Dirichlet process mixture}
\acrodef{TOA}[TOA]{time of arrival}
\acrodef{GMM}[GMM]{Gaussian mixture model}
\acrodef{OOK}[OOK]{on-off keying}
\acrodef{MLE}[MLE]{maximum likelihood estimation}
\acrodef{LOS}[LOS]{line-of-sight}
\acrodef{NLOS}[NLOS]{non-line-of-sight}
\acrodef{SG}[SG]{signal generator}
\acrodef{SEP}[SEP]{Sun-Earth-probe}
\acrodef{FDSOI}[FDSOI]{fully depleted silicon on insulator}
\acrodef{OpEx}[OpEx]{operational expenditures}
\acrodef{TCO}[TCO]{total cost of ownership}
\acrodef{CapEx}[CapEx]{capital expenditures}
\acrodef{MAC}[MAC]{medium access control}
\acrodef{GEO}[GEO]{geostationary orbit}
\acrodef{SWaP}[SWaP]{size, weight, and power}
\acrodef{NOMA}[NOMA]{Non-orthogonal multiple access}
\ifCLASSINFOpdf
\else
\fi

\begin{document}
\title{Reconfigurable Intelligent Surfaces in Action for Non-Terrestrial Networks}

\author{K{\"{u}}r{\c{s}}at~Tekb{\i}y{\i}k,~\IEEEmembership{Graduate Student Member,~IEEE,} G{\"{u}}ne{\c{s}}~Karabulut~Kurt,~\IEEEmembership{Senior~Member,~IEEE,} Ali~Rıza~Ekti,~\IEEEmembership{Senior~Member,~IEEE,} Halim~Yanikomeroglu,~\IEEEmembership{Fellow,~IEEE}

\thanks{K. Tekb{\i}y{\i}k is with the Department of Electronics and Communications Engineering, {\.{I}}stanbul Technical University, {\.{I}}stanbul, Turkey. e-mail: tekbiyik@itu.edu.tr}

\thanks{G. Karabulut Kurt is with the Department of Electrical Engineering, Polytechnique Montr\'eal, Montr\'eal, Canada, e-mail: gunes.kurt@polymtl.ca}

\thanks{A.R. Ekti is with the Grid Communications and Security Group, Electrification and Energy Infrastructure Division, Oak Ridge National Laboratory, Oak Ridge, TN, U.S.A., e-mail: ektia@ornl.gov}

\thanks{H. Yanikomeroglu is with the Department of Systems and Computer Engineering, Carleton University, Ottawa, Canada. e-mail: halim@sce.carleton.ca}
 
}

\IEEEoverridecommandlockouts 

\maketitle

\begin{abstract}
Next-generation communication technology will be made possible by cooperation between terrestrial networks with \ac{NTN} comprised of high-altitude platform stations and satellites. Further, as humanity embarks on the long road to establish
new habitats on other planets, the cooperation between \ac{NTN} and \ac{DSN} will be necessary. In this regard, we propose the use of reconfigurable intelligent surfaces (RIS) to improve coordination between these networks given that RIS perfectly match the \acl{SWaP} restrictions of operating in space. A comprehensive framework of \acs{RIS}-assisted non-terrestrial and interplanetary communications is presented that pinpoints challenges, use cases, and open issues. Furthermore, the performance of \acs{RIS}-assisted \ac{NTN} under environmental effects such as solar scintillation and satellite drag is discussed in light of simulation results.
\end{abstract}

\begin{IEEEkeywords}
	Reconfigurable intelligent surfaces, non-terrestrial networks, deep-space networks.
\end{IEEEkeywords}

\IEEEpeerreviewmaketitle
\acresetall

\section{Introduction}\label{sec:intro}

\begin{figure*}[!t]
    \centering
    \includegraphics[width=\linewidth, page=1]{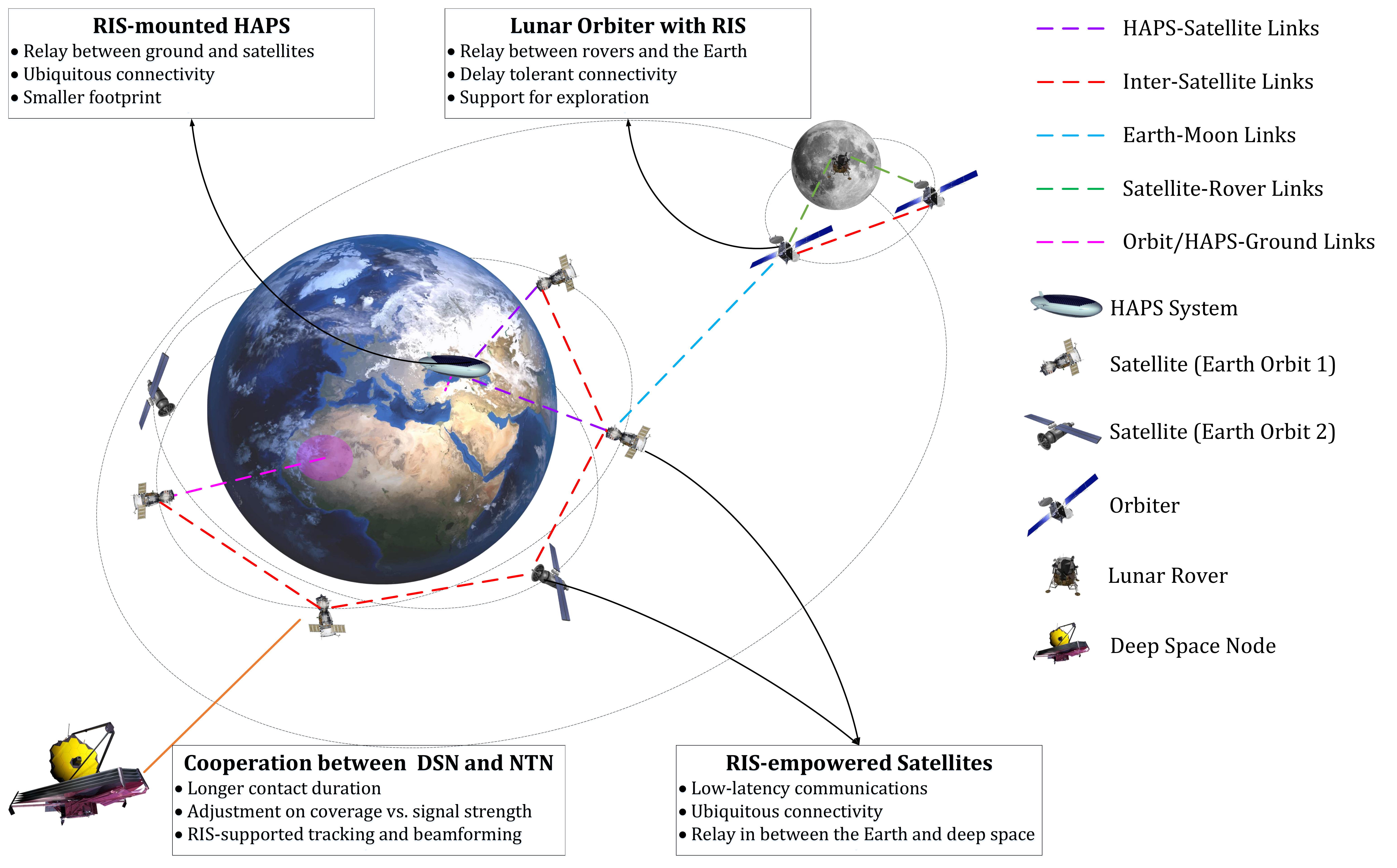}
    \caption{RIS can enable low-cost and energy-efficient non-terrestrial communication links including HAPS, satellites, and even deep-space communications nodes such as the lunar relays and space telescopes. Furthermore, the mechanical conformity of RIS makes light-weight and high-capacity communication subsystems in satellites possible.}
    \label{fig:general_view}
\end{figure*}

It is widely accepted that user-centric and ubiquitous connectivity, which are desired by both end-users and operators for 6G and beyond, can be achieved through unique orchestration of terrestrial and \ac{NTN} in next-generation communication systems~\cite{tekbiyik2020holistic}. This vision is also described by the 3rd Generation Partnership Project (3GPP) in TR~38.821 for the operation of New Radio (NR) in \ac{NTN}. According to the definition by 3GPP, an \ac{NTN} basically consists of unmanned aerial vehicles, \ac{HAPS} systems, and satellite deployments. \ac{LEO} satellites and \ac{HAPS} systems are considered to be the key enablers for \ac{NTN} due to their unique features, which include longer operating times and wider coverage areas~\cite{kurt2020vision}. Thus, this study focuses on HAPSs and LEO satellites in the context of NTN. HAPS systems, positioned at altitudes of $20-50$ km possibly in a quasi-stationary manner, can provide significant advantages while connecting space and terrestrial networks~\cite{kurt2020vision}. LEO satellites, for their part, present a remarkable potential for \ac{NTN}, serving as a densely deployed near-Earth network to enable connectivity to and from the Earth.

The number of space missions conducted by various organizations led by NASA is increasing. It is aimed that humanity will again reach the Moon by starting space travels in 2024. There has also been an increase in studies of Mars, and ways to establish a second world for humanity are being sought. For instance, a relay network composed of multiple orbiters was used during the landing of Perseverance and the transmission of data on the red planet, which is a step that NASA has recently carried out as part of its Mars mission. Although the communication systems installed in space for this purpose are advanced technology, their service quality is far below that of terrestrial communication technology due to limitations for \ac{SWaP} and cost associated with operating in deep-space. For example, the link between Perseverance and the Mars orbiter can only provide a data rate of $2$ Mbps~\cite{Communic97online}. On the other hand, several space missions have been planned to explore outer space. The \ac{DSN} are one of the key elements for long-range exploration. \ac{DSN} can be introduced as a network supporting interplanetary and Earth-orbiting missions\footnote{It should be noted that there is no consensus on the definition of DSN. Unlike ITU, NASA keeps the scope of the definition broader to cover the networks supporting and Earth-orbiting missions as well as interplanetary missions. In this paper, we follow NASA's definition except for that satellites in Earth's orbits for DSN.}.

As illustrated in \FGR{fig:general_view}, \ac{LEO} satellites and \ac{HAPS} systems have the potential to operate as key relays between \ac{DSN} and terrestrial networks due to the seamless and ubiquitous connectivity they provide. In the next years, data traffic both in HAPS systems and LEO networks is expected to increase because of proliferation of satellites and high data demand by space missions. However, due to the increased transmission distances, the transmission power required will also increase to compensate for the high path loss. As \ac{HAPS} systems and \ac{LEO} satellites are strictly limited by the \ac{SWaP} requirements, they can employ fewer RF chains. \ac{RIS}, which have been recently proposed to maximize the \ac{SNR} at the receiver by adjusting the incident wave phase with only a single RF chain~\cite{huang2019reconfigurable}, can be employed in \ac{NTN} to improve energy efficiency~\cite{giordani2020non}. Due to drawbacks with current antenna array systems in both NTN and DSN, RIS could improve communication performance in a passive and energy-efficient way. The seminal works (e.g.,~\cite{basar2019wireless, huang2019reconfigurable}) on RIS demonstrate that they can offer better communication performance while consuming less power.

Since geostationary orbit (GEO) satellites are fixed in position relative to ground stations, beamforming and beam tracking are not as essential as LEO satellites. But, RIS can also be utilized in GEO satellites to increase received signal strength rather than beamforming. On the other hand, medium Earth orbit (MEO) satellites have a longer observation time from a particular location on the ground. So, it avoids frequent handover compared to LEO satellites. RIS can be employed in MEO constellations for both passive beamforming and energy efficiency.

By leveraging the beamforming and steering capabilities of RIS, it is possible to track near-Earth satellites and orbiters for a longer time. To do this, reflect-arrays and phased-arrays are currently used in deep-space systems. Although reflect-arrays can produce a sharp beam, they cannot steer. Besides, phased-arrays suffer from both cost and \ac{SWaP} drawbacks as they contain multiple active RF elements. The main differences between them are summarized in~\TAB{tab:summary}. RIS therefore appear to be a promising solution for both NTN and DSN, since RIS include only a single RF chain, reduces the cost and size. This study addresses possible use-cases and challenges of \ac{RIS} in \ac{NTN} and \ac{DSN} considering the unique characteristics of these networks while highlighting the potential of \ac{DL} techniques to address challenging channel environments.

The rest of the paper is organized as follows. In~\SEC{sec:ris_motivation}, we discuss the motivation behind RIS in non-terrrestrial and deep-space communications. \SEC{sec:use_cases} introduces possible use cases for \ac{RIS}-assisted \ac{NTN}. \SEC{sec:challenges} lists the challenges and introduces possible solutions, which are supported with numerical results. Open issues for the realization of \ac{RIS}-assisted \ac{NTN} and \ac{DSN} are discussed in~\SEC{sec:open_issues}. \SEC{sec:conclusion} concludes the study.

\begin{table}[!t]
\centering
\caption{The main differences between RIS, reflect-arrays, and phased-arrays.}
\begin{tabular}{cccc}
\toprule \toprule

\textbf{}        & \textbf{RIS} & \textbf{Reflect-arrays} & \textbf{Phased-arrays} \\ \toprule
\textbf{Beam Steering}                & +                                & --                                           & +                                          \\
\textbf{Adaptive Beamforming}         & +                                & --                                           & +                                          \\
\textbf{Hardware Complexity}          & Low                              & Low                                         & High                                       \\
\textbf{Components}             & Passive                           & Passive                                         & Active \\
\textbf{Channel Estimation} & High                             & High                                        & Medium                                     \\
\textbf{Number of RF Chains}          & 1                                & 1                                           & \textgreater{}1                            \\
\textbf{Active Components}            & No                               & No                                          & Yes                                        \\
\textbf{Weight}                       & Low                              & Low                                         & High                                       \\
\textbf{Fabrication Cost}             & Medium                           & Low                                         & High \\
\bottomrule \bottomrule
\end{tabular}
\label{tab:summary}
\end{table}

% \newpage
\section{Why \acl{RIS}s?}\label{sec:ris_motivation}

In addition to being a candidate technology for 6G, \ac{RIS} communications present a remarkable potential for future wireless communications. There are numerous attractive features of utilizing \ac{RIS} in \ac{NTN}, but three of them stand out: energy efficiency, compatibility with \ac{SWaP} requirements, and affordability~\cite{basar2019wireless}. 

\subsection{Energy Efficient System Architecture}
 
While \ac{HAPS} and satellite systems are expected to yield significant gains in terms of capacity and coverage for wireless communications, we should not forget that these systems have energy constraints~\cite{kurt2020vision}. Therefore, it is necessary to carefully consider energy management in system designs. Along this line, solar-powered \ac{HAPS} systems can be utilized for wireless communications, but this does not mean that the energy of a HAPS or satellite is unlimited. The capacity of fuel cells is one of the lead determinants of a satellite's operational life. It should be noted that the ability of \ac{RIS} to configure the phases of meta-atoms to maximize the \ac{SNR} at the receiver can provide a considerable amount of gain for an energy-efficient communication system. 

Owing to the virtual \ac{MIMO} structure and single RF chain of RIS, when they are used as transmitters, they require less operational and transmission energy than traditional \ac{MIMO} systems that contain many active circuit elements~\cite{basar2019wireless}. In the context of Lunar relay systems and other relays in DSN, \ac{RIS} can enable low-power relay systems in deep-space due to their capacity to maintain almost the same \ac{EIRP} with a significantly reduced power consumption~\cite{huang2019reconfigurable}.

\subsection{Compatibility with SWaP Requirements}\label{sec:swap}

Besides the production cost of satellites, the cost of launching satellites into orbit is also quite high. One of the main factors affecting this cost is the weight of a satellite system. Although space transport today costs less than in the past, launches still require satellite operators to have a significant budget. To illustrate, the average launch between the years 1970 and 2000 cost $\$16000$ per kg but this has decreased to $\$6000$ per kg as of 2018~\cite{jones2018recent}. Recently proposed reusable launch systems have reduced the cost of space missions, but it is still quite high. A reduction in the weight of satellites and \ac{HAPS} systems can reduce the cost per operation significantly. Furthermore, aircraft and spacecraft (e.g., \ac{HAPS}, satellites, orbiters) can carry a limited payload due to their aerodynamics. 

\ac{RIS} can be smaller in size than phased array structures while still providing extensive coverage~\cite{huang2020holographic}. In order to serve ground users with small antennas, satellite antennas are designed to be as large as possible. This situation increases the risk of collisions due to the increasing density of satellite deployments. The main factors in the deployment cost of a \ac{HAPS} are weight, power consumption, and flight duration. Utilizing \ac{RIS} can reduce operational costs and extend flight duration of \acp{HAPS}, since \ac{RIS} are both lightweight and low-power consuming devices compared to traditional phased-arrays. It is expected that the costs can gradually reduce by utilizing \ac{RIS} in \ac{NTN}.

Given that \ac{HAPS} systems and satellites are the first point of contact between Earth and deep-space, they can be considered to increase the signal power received or transmitted by using \ac{RIS} which enhance the \ac{QoS} and energy efficiency.

\subsection{Relatively Affordable Alternative Solution}

In addition to the lower operational costs discussed above, another important factor that makes \ac{RIS} attractive is that they can be produced at a relatively lower cost compared to conventional multi-antenna systems. The main reason for this is that \ac{RIS} include a single RF chain which is the most costly part of the transceivers. For example, it has been reported that the first prototype introduced in~\cite{dai2020reconfigurable} reduced hardware costs compared to those of phased array systems.

\section{Use Cases}\label{sec:use_cases}
Although \ac{NTN} standardization is underway by 3GPP in Rel.~16 in TR~38.821, the use cases extend well beyond densely deployed mega-constellations. This section focuses on RIS use cases in high-altitude systems and space networks.

\paragraph{\ac{RIS}-assisted \ac{HAPS}}
Here, we consider scenarios involving RIS on HAPS systems. Ultra-dense deployment of \ac{HAPS} systems can enhance ground coverage and eliminate coverage gaps~\cite{kurt2020vision}. Furthermore, it is possible to provide seamless connectivity by handling the handoffs of \ac{LEO} satellites. By making use of the large surfaces of a \ac{HAPS} for the placement of \ac{RIS}, both coverage and energy efficiency can be improved. In addition to serving rural areas, HAPS systems can also increase the overall throughput of the entire network. In addition, as the payload weight and the amount of energy consumed by the communication payload decrease with the use of \ac{RIS}, the average flight duration of a \ac{HAPS} is expected to increase. 

By enabling \ac{RIS}-assisted \ac{HAPS}-satellite integrated networks, a multi-\ac{RIS} scheme can be used to increase the error performance and achievable data rate. Furthermore, the prediction that dense \ac{LEO} satellite deployments will be an important enabler of next-generation communication systems increases interest in investing in space technology. The use of \ac{RIS} in \ac{LEO} satellite networks shows that both error performance and achievable data rates can be significantly improved~\cite{tekbiyik2020reconfigurable}. Since it is expected that \ac{HAPS} systems will bridge the space and terrestrial networks, outfitting them with \ac{RIS} introduces a significant performance advantage.

\paragraph{Integration with \ac{DSN} towards Human Settlement in Space} 

Communication and navigation systems will be indispensable subsystems in future space missions. For instance, NASA plans to land on the Moon's south pole by 2024 as part of the Artemis project. This landing is connected with the broader objective of building permanent structures, settling, and ultimately sending humans to Mars. However, some communication and navigation issues must be handled carefully since space agencies now aim to establish a living station like a Lunar Gateway. Thus, a reliable and secure communication link between a control station on Earth and a Lunar Gateway will be indispensable.

Furthermore, this communication system will need to support high data rates in order to transmit high-resolution camera and sensor data for mapping celestial bodies and analyzing soil in interplanetary reconnaissance. In terms of manned reconnaissance, it can be clearly predicted that communication security and continuity will be vital. For example, it takes up to $20$ hours for NASA to receive $250$ megabits of data transmitted directly from Mars to the Earth~\cite{mars_comm}. The main reasons for this are that the rover can be in \ac{LOS} with the Earth for only a few hours a day, and the rover's transmission power is limited. By utilizing \ac{RIS} with non-terrestrial (i.e., \ac{DSN}), the transmission time can be extended by employing multi-hop relays with steering capabilities. When a direct LOS is blocked, it will still be possible to transmit information over consecutive multi-\ac{RIS} satellites and improve error performance, as shown in~\cite{tekbiyik2020reconfigurable}. Also, a rover can direct beams towards a target by utilizing passive beamforming supported by \ac{RIS}. Likewise, Perseverance uses a high gain steerable antenna for the same aim~\cite{Communic97online}.

A \ac{RIS}-mounted relay station can be deployed on Mars to extend the \ac{LOS} duration by leveraging the \ac{RIS}'s beamforming capabilities. The reflect-arrays have currently been utilized in \ac{DSN}. Without essential modifications on the system, the performance can be improved by only adjusting the phase shifts~\cite{basar2019wireless}. By providing continuous connectivity, \ac{IP}-based networks would start to take the place of \acp{DTN} operating with the store-and-forward paradigm. Especially in cases where remotely controlled vehicles and rovers are employed during the operation, this type of communication protocol is very likely to cause failure of the operations. Supporting extra-terrestrial cellular communication systems with \acp{V-HetNet}~\cite{tekbiyik2020holistic} together with RIS would be a game-changer for space communications. Also, Cubesats deployed in LEO have flourished in the recent space missions. However, using them in deep-space missions requires higher transmit power or high gain antennas~\cite{chahat2019advanced}. As the stowage and power capacity of Cubesats are limited, relays might be used in between NTN and DSN. Long-range communication with Cubesats can be enabled by RIS-assisted relay satellites that provide cooperation between NTN and DSN units.

\section{Operational Challenges}\label{sec:challenges}

Although RIS have many attractive features, they are still at an early stage of being integrated into non-terrestrial systems and there are many challenging tasks still to be addressed. Due to long-distance and short durations of scheduled direct communication, the transmitted data cannot be received properly via \ac{IP}. A physical layer boost involving an \ac{SNR} improvement will certainly be useful to the connectivity of these links. In this section, we focus on the challenges related to this physical layer, which are expected to have a significant effect on the link performance, and we also numerically highlight the potential of using \ac{RIS} to increase the \ac{SNR}.

\subsection{\ac{RIS} Deployment on NTN Systems}
The first and most fundamental problem in implementing RIS-powered aerial systems is undoubtedly the integration of smart surfaces in aerial platforms. Considering the flight aerodynamics and mechanical structures of a \ac{HAPS} or satellites, it is clear that a serious design problem needs to be addressed: namely that depending on changes in position and elevation angles some or all meta-atoms are likely to be shaded by the system itself.

Conformal metasurfaces can facilitate RIS deployments since conformal metasurfaces can allow the coating of irregular surfaces or arbitrarily shaped 3-D objects. Thus, allowing the coating of spherical and elliptical surfaces with \ac{RIS}. By utilizing artificial impedance surfaces, the generation of any desired radiation pattern becomes possible when the surface is designed to integrate antennas into complex-shaped objects. An \ac{RIS} can be deployed on the surface of a \ac{HAPS} and solar panels of relay-satellites providing the cooperation between NTN and DSN. Also, \ac{RIS} can also be used to replace the antenna subsystems already on satellites. It should be noted that different application-specific \ac{RIS} designs may be needed in terms of material and conformal coating. Moreover, using lightweight material would be important in the RIS design to reduce launch costs. The reflection loss, the number of phase values RIS can support, and the weight will be an important design and optimization problem in RIS design. To comply with rockets' size constraints, the foldable RIS can be a key player in DSN.

\subsection{Solar Scintillation}\label{sec:scintillation}

\begin{figure}[!t]
    \centering
    \includegraphics[width=\linewidth, page=2]{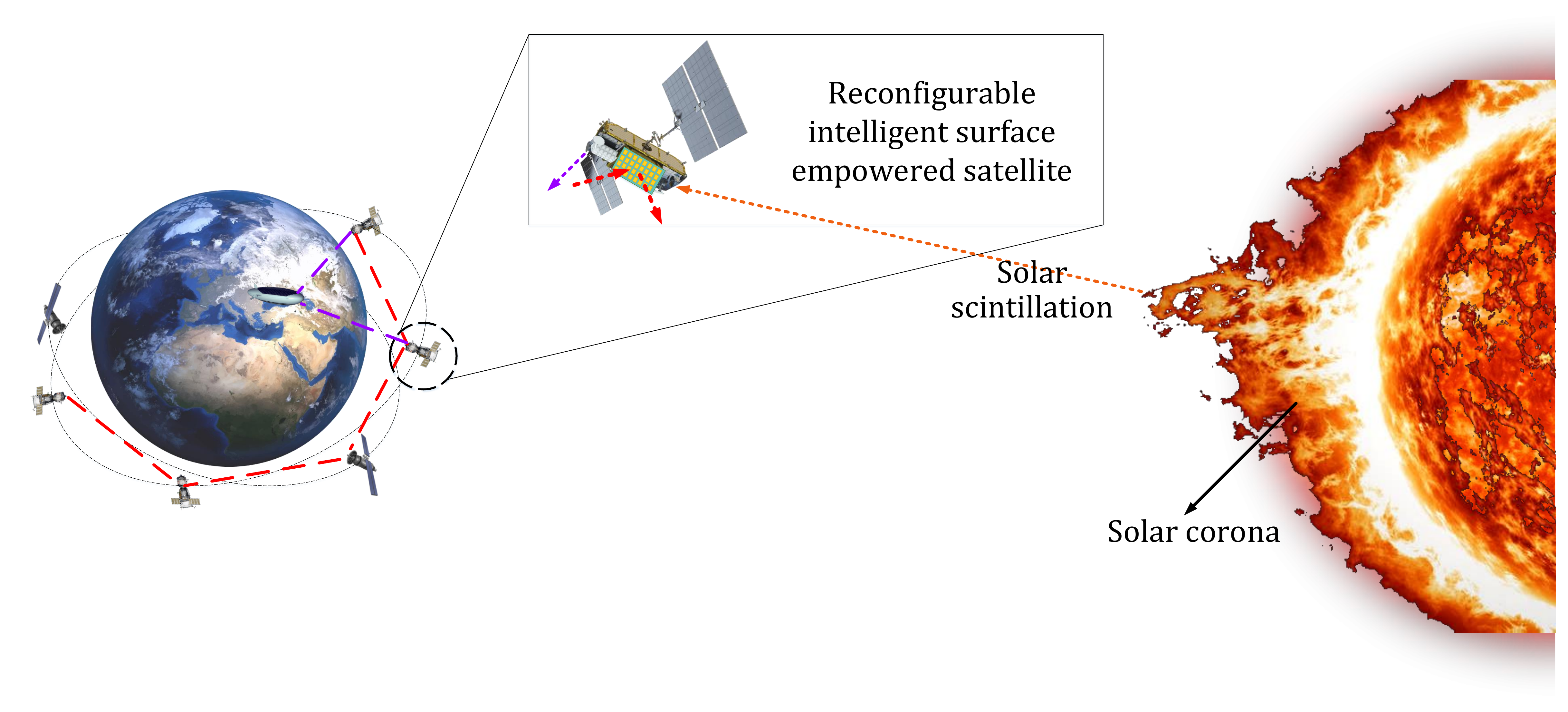}
    \caption{Solar scintillation is a phenomenon caused by changes in plasma density in solar winds ejected from the Sun. Solar scintillation crucially degrades the performance of space communications by affecting the power and phase of the transmitted signal due to diffraction and refraction.}
    \label{fig:solar_scintillation}
\end{figure}

\begin{figure}[!t]
    \centering
    \includegraphics[width=\linewidth]{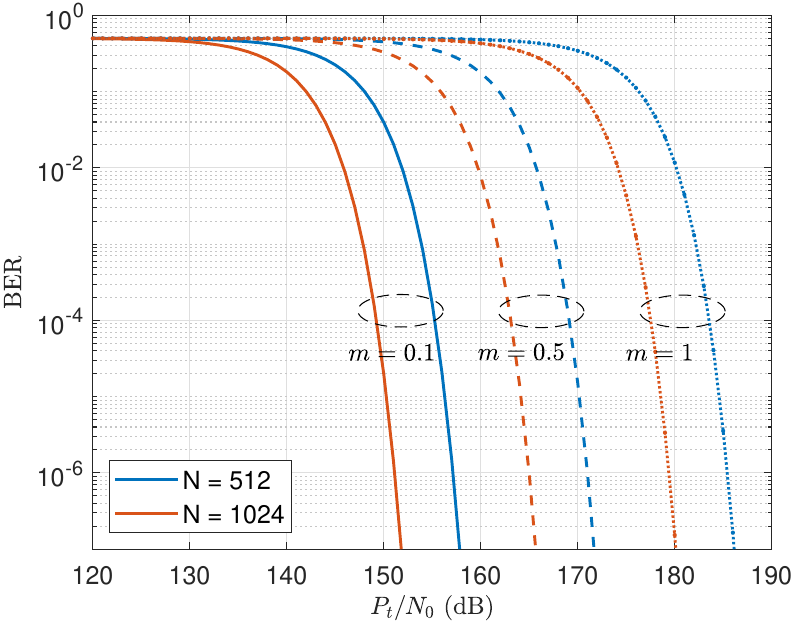}
    \caption{\ac{BER} performance analysis of an \ac{RIS}-assisted \acl{ISL} of a Starlink-like constellation whose inter-satellite distance is $945.4$ km, which indicates that the solar scintillation can be diminished by increasing the number of meta-atoms. The number of meta-atoms, scintillation index, and transmit power-to-noise ratio are denoted by $N$, $m$, and $P_{t}/N_{0}$, respectively.}
    \label{fig:ber_scintillation}
\end{figure}

It is vital to establish robust and reliable communication links between the Earth and remote space stations, reconnaissance robots, and even astronauts for deep-space missions. In order to provide reliable communications, the space environment should be accurately modeled. It should be noted that many factors --some of which are not yet known-- may disrupt the structure of the transmitted signals in space.

Solar scintillation, is caused by irregularities in plasma density in solar winds as illustrated in~\FGR{fig:solar_scintillation}. Solar scintillation substantially decreases communication performance. The scintillation effect reaches its peak during the solar conjunction, which refers to the alignment of the Earth, Sun, and another communication node on the same line. The \ac{SEP} angle can be considered as a metric to denote how close celestial bodies are to the solar conjunction.

To illustrate, the SEP angle between the Earth and Mars is below $20^{\circ}$ for $170$ days over a $780$-day period~\cite{xu2019effects}. This means that solar scintillation would have a serious effect for much of this $780$-day period. To address this, we investigate the impact of solar scintillation on an \ac{RIS}-assisted \ac{ISL} with \ac{BPSK} signaling. In accordance with mmWave frequencies allocated for \acp{ISL} by ITU-R S.159, the link between two satellites in the same orbit is simulated under the assumption of the Starlink constellation at $23$ GHz. The distance between two nearby satellites is approximately $945.4$ km. It is therefore assumed that the satellites are operating under solar scintillations with varying strengths. We refer the readers to~\cite{tekbiyik2020reconfigurable} for the rest of simulation parameters and details. The \ac{BER} performance of the \ac{RIS}-assisted intra-plane \ac{ISL} is depicted in~\FGR{fig:ber_scintillation}. The error performance is inversely proportional to the square of the scintillation index at a low scintillation levels~\cite{tekbiyik2020reconfigurable}. Compared to a weak scintillation effect, it is necessary to increase the output power by approximately $14$ dB in order to remain the same error performance in the transition zone. These results reveal that communication units should adapt their transmit power according to their positions relative to the Sun.

\subsection{Misalignment Fading}

\begin{figure}[!t]
    \centering
    \includegraphics[width=0.9\linewidth, page=3]{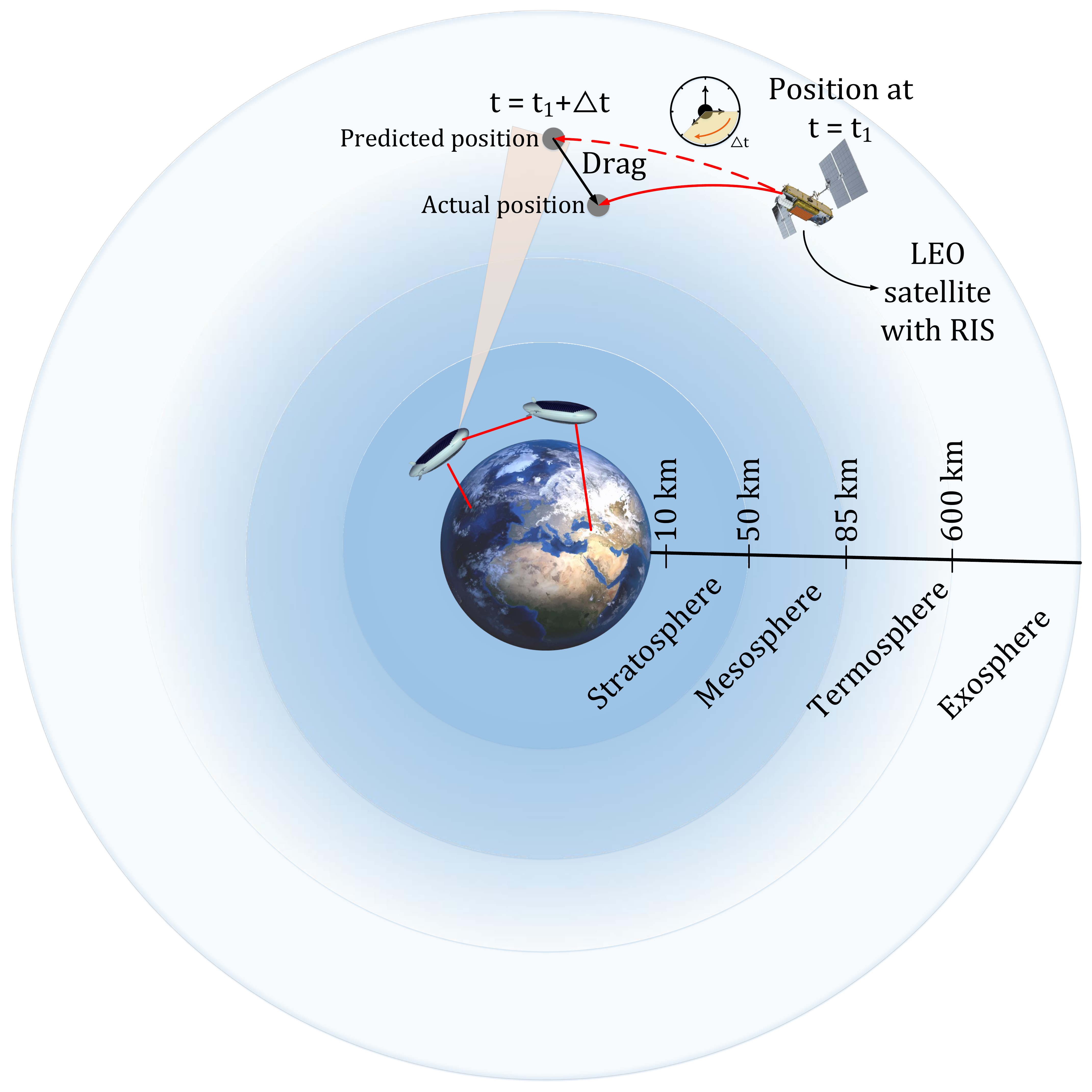}
    \caption{Atmospheric drag results in the satellite moving towards the Earth and can destabilize cooperation between HAPS deployment and LEO satellites over time. The satellite, which is in orbit at time $t_1$, may have lost its altitude at $t=t_1+\Delta t$ due to the atmospheric drag. Due to movement, it is possible to observe a misalignment between transceivers.}
    \label{fig:atmospheric_drag}
\end{figure}

Another challenging aspect of \ac{RIS}-assisted \ac{NTN} is the misalignment fading that can occur due to an error in the alignment between the transmitter and receiver antennas. There may be various reasons for this error. For instance, the position and altitude of a HAPS system in the Earth's stratosphere might change suddenly on account of heavy wind and related atmospheric disturbances. Also, the density of gases in the mesosphere is high, so there is likely to be molecular absorption and the refraction of waves. This layer of the atmosphere is warmer than the others, as most of the radiation from the sun is absorbed by the particles in the thermosphere. The difference in density between the layers may cause diffraction of the waves, which may cause misalignment. Furthermore, heating and expanding in the atmosphere due to the particles, X-ray, and ultraviolet beams ejected from the Sun creates an irregularity in the density of atmospheric layers. Then, a drag force, which is called atmospheric drag, is observed owing to the flow from higher density layers to lower ones. This drag force can cause a satellite to lose its altitude as illustrated in~\FGR{fig:atmospheric_drag}, which is named satellite drag.

In order to ensure reliable communications via \ac{NTN}, it is critical to evaluate possible misalignment effects on RIS-assisted communication systems. In our previous work~\cite{tekbiyik2020reconfigurable}, misalignment fading denoted by jitter, $\sigma_{s}$, was considered in \ac{RIS}-aided communications for \acp{ISL}. Drawing on the findings in~\cite{tekbiyik2020reconfigurable}, we examine the BER performance of RIS-assisted intra-plane ISLs with misalignment fading by employing the same simulation parameters as in \SEC{sec:scintillation}. The rest of simulation parameters is detailed in~\cite{tekbiyik2020reconfigurable}. \FGR{fig:ber_misalignment} demonstrates that significantly high power is needed to maintain the same BER performance level. $250$ dB of additional power is required to keep the \ac{BER} at a level of about $10^{-7}$ when the offset between the centers of the receiver and transmitter antenna beams is around $1$ m. The simulation results regarding misalignment fading in \ac{RIS}-assisted communication systems show the necessity for instantaneous beam alignment methodologies, which are still open in the literature.

\begin{figure}[!t]
    \centering
    \includegraphics[width=\linewidth]{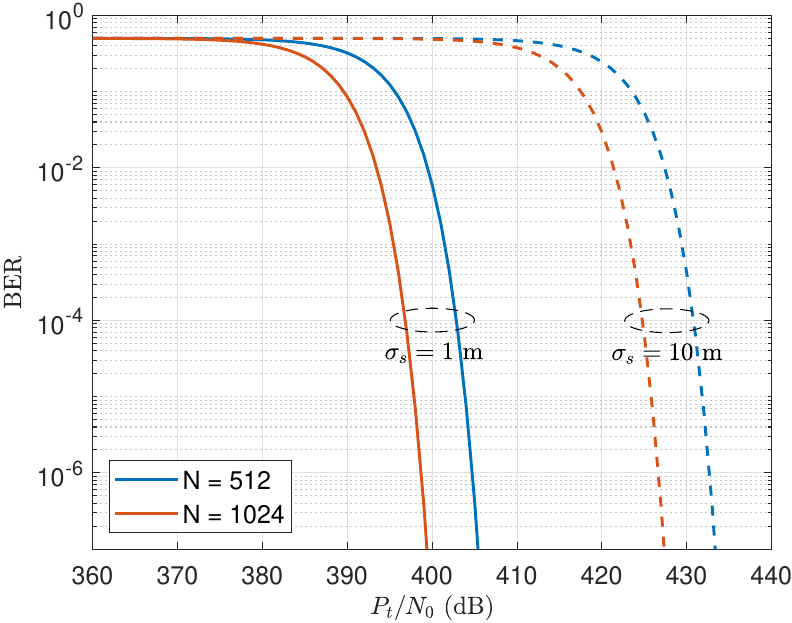}
    \caption{\ac{BER} versus $P_{t}/N_{0}$ under the misalignment fading, which can seriously decrease the communication performance in terms of error probability, for \ac{BPSK} signaling. The radius of equivalent beamwidth is chosen as $10$ m and $\sigma_{s}$ denotes the misalignment jitter.}
    \label{fig:ber_misalignment}
\end{figure}

\subsection{Channel Estimation}

Because \ac{RIS} can adjust the amplitude and/or phase of the incident electromagnetic wave, they can largely override the randomness of the propagation medium. But doing this certainly requires high-quality \ac{CSI} first. Therefore, channel estimation plays a vital role in terms of \ac{QoS} in non-terrestrial communication as well. For this purpose, \ac{DL} can be used to achieve high performance in channel estimation under challenging channel conditions. Recently, we proposed the use of \acp{GAT} for channel estimation in \ac{RIS}-assisted full-duplex \ac{HAPS} backhauling~\cite{tekbiyik2020channel}. This model can also be used in different network architectures. The \ac{GAT} can reduce the computational complexity and increase the learning capacity by taking unseen nodes within the proposed \ac{RIS}-assisted \ac{NTN} in the process~\cite{tekbiyik2020channel}. The model is trained by assigning the received signal samples and known pilot samples to the nodes and vertices of graphs, respectively. By utilizing the trained model proposed in~\cite{tekbiyik2020channel}, we investigate the channel estimation performance for \acp{ISL} between two neighboring satellites in the same orbit under varying solar scintillation. \FGR{fig:nmse_scintillation} indicates that the proposed model is robust against the changes in the channel characteristics. Not even strong solar scintillation degrades the performance of the \ac{GAT} estimator trained over a dataset that does not include solar scintillations. 

The proposed model can also estimate channel coefficients with almost the same high performance as at various solar scintillation levels, namely weak, transition zone, and strong. The reason for this performance is the attention mechanism which enables the model to focus on the most relevant information in the input; hence, \ac{GAT} can be effective under changing channel parameters even if the model has been trained under better conditions than those of the test stage. It can be concluded that GATs shows a high performance in RIS-supported cooperation of NTN and DSN despite changing channel conditions. However, further research along this direction is required for a holistic exploration.

\begin{figure}[!t]
    \centering
    \includegraphics[width=\linewidth]{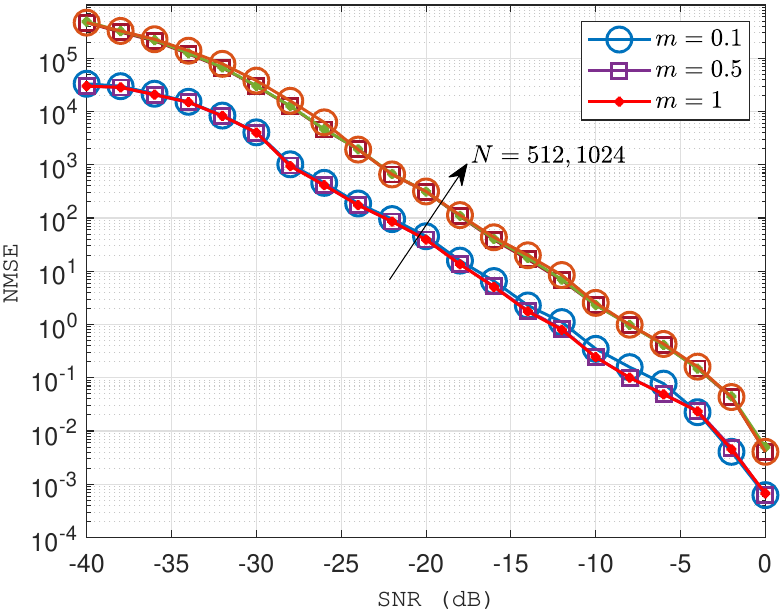}
    \caption{\ac{GAT} can perform accurate channel estimation under varying solar scintillation effects, which can be grouped into weak ($m<0.3$), transition zone ($0.3\leq m<1$), and strong scintillation ($m=1$). The circular, square and dot markers denote misalignment index $m = 0.1$, $0.5$, and $1$, respectively.}
    \label{fig:nmse_scintillation}
\end{figure}

\vspace{-0.2cm}
\section{Open Issues and Research Directions}\label{sec:open_issues}

\ac{RIS}-assisted \ac{NTN} have the potential to significantly improve the capacity and reliability of personal communications and deep-space communications. Despite these promises, many open issues remain and must be carefully considered.

\paragraph{Channel Models for \ac{NTN} and \ac{DSN}}

As is well known, an understanding of channel behavior on both a large scale and small scale is essential for developing a link budget analysis and determining the waveform and its parameters. However, the \ac{RIS}-assisted network topologies discussed in this study are challenging in two ways. 

First, channel models for stratospheric transceivers do not allow a proper understanding of the channel behavior as the atmospheric effects have not yet been considered in detail~\cite{kurt2020vision}. Also, \ac{HAPS}-to-\ac{HAPS} channels have yet to be investigated for mobility cases. 3-D deployments are now replacing 2-D deployments, channel models need to be handled in accordance with this dimensional expansion. Furthermore, channels between satellites and \ac{HAPS} systems should be examined in terms of both mobility and near-space effects. 

Conversely, since channel modeling studies for satellites and \ac{DSN} are still at an early stage, this area also needs to be developed with appropriate channel models. For example, while the small-scale fading effect created by the solar scintillation in the amplitude of the signal is Rician distributed for low scintillation levels, a full-fledged model has not yet been proposed for strong scintillations~\cite{xu2019effects}. Therefore, the coronal channel modeling should be further studied. A comprehensive study on \ac{DSN} and channel models suitable for the atmospheres of other planets is also essential.

The second challenge is that all these channel modeling studies should also be generalized for \ac{RIS}-assisted communication, but RIS channel models are still at an early stage, even for terrestrial use cases. Although meta-atoms are referred to as theoretically perfect reflectors, their performance may vary depending on the type of material used in practice. Also, the correlation between meta-atoms can affect the behavior of the communication channel~\cite{basar2019wireless}. For this reason, the channel models recommended for \ac{RIS} must first be verified through measurements.

\paragraph{Design and Operation of \ac{RIS} under Harsh Environmental Conditions} 

Temperature changes and other environmental factors must be taken into consideration in the design of RIS. RIS must be designed in such a way that they are resistant to temperature differences between day and night. Even though space is presumptively accepted as a vacuum, solar winds fill space with plasma and particles ejected from the Sun. Furthermore, coronal mass ejections can impact wireless channel by changing the radiation and temperature in the operating environment of the \ac{RIS}-assisted nodes. For example, charged particles in Van Allen Belts can damage the electronic components of communication systems. Thus, radiation-tolerant \ac{RIS} and all instrumentation making \ac{RIS} operational must be investigated. Dust storms on other planets and temperature differences occurring over a short time duration are also important considerations.

Furthermore, considering the current narrow-band designs, they might not fully meet the demands of future wideband ISLs. Microfluidically reconfigurable intelligent surfaces (MRIS) can support wideband applications~\cite{entesari2016fluidics}. Also, their inherent continuous phase shift can further improve communication performance. But, they are not expected to perform well in applications that require frequent reconfiguration due to relatively slow switching in MRIS.

\paragraph{Super Smart Energy Management}

To be able to manage and maintain a high performance \ac{NTN}, smart and efficient energy management should be investigated. Active use of energy harvesting from hybrid sources, including RF and solar energy, is a necessity in resource-limited deployment scenarios. 
Distributed low-power sensor networks in planetary exploration can pave the way towards an Internet of Space Things, which will result in a prolific amount of information for deep-space. These sensor networks can be powered by harvesting RF-energy with RIS. Also, wearable biomedical devices can harvest energy via \ac{RIS}, as continually monitoring the health status of astronauts is essential. Moreover, the potential benefits of \ac{RIS} in transmitting power from solar power satellites to a planetary surface can be considered.

Managing this extremely complex multi-connectivity network with multi-layer communication by integrating \ac{NTN} with \ac{DSN} can be supported by machine learning algorithms. Conventional radio resource and energy management methods might be inadequate to provide the desired \ac{QoS}. Reinforcement learning both leverages communication performance and provides the minimum power consumption by solving complex optimization problems with near-optimal results.

\paragraph{Performance Analysis}

Although this study addresses BER and NMSE in RIS-assisted satellite networks, RIS-assisted NTN and DSN should be considered in terms of several performance metrics. Compared to terrestrial applications, energy efficiency and received signal strength are much more important in non-terrestrial applications due to long-distance communication. Therefore, further research on the energy efficiency of RIS is required. Furthermore, to the authors' best knowledge, there is no study yet that addresses the handoff rate in RIS-powered mega-constellation satellite networks. Considering the motion of space-things, handoff rate is the candidate to be a crucial performance metric for RIS-assisted NTN and DSN. Moreover, the handoff rate for RIS-assisted LEO and MEO satellites can be compared to optimize the satellite constellations.

\section{Concluding Remarks}\label{sec:conclusion}

In this paper we proposed an RIS-assisted DSN integrated with NTN to extend the duration of active links between Earth and space stations by using relay stations compatible with SWaP requirements. The challenges and opportunities were detailed by interpreting simulation results. 
Open issues and research directions for using \ac{RIS} in \ac{DSN}-integrated \ac{NTN} were elaborated. Perhaps in the near future the sound of ``one giant leap for mankind'' will reach us via \ac{RIS}-assisted \ac{DSN} integrated with \ac{NTN}—but with better quality and reliability, of course.

\balance

\bibliographystyle{IEEEtran}
\bibliography{main}

%\vspace{-0.2cm}

%%%%%%%%%%%%%%%%% BIOGRAPHIES %%%%%%%%%%%%%%%%%%%%%%
\section*{Biographies}
\footnotesize{K{\"{U}}R{\c{Ş}}AT TEKBIYIK [StM'19] (tekbiyik@itu.edu.tr) is pursuing his Ph.D. degree in Telecommunications Engineering at Istanbul Technical University and is also Wireless Systems and Machine Learning Engineer at Assia Inc.\\

G{\"{U}}NE{\c{Ş}} KARABULUT KURT [StM'00, M'06, SM'15] (gunes.kurt@polymtl.ca) received her Ph.D. degree in electrical engineering from the University of Ottawa, Ottawa, ON, Canada, in 2006. She is with Polytechnique Montréal. She is also an adjunct research professor at Carleton University. Gunes is currently serving as an Associate Technical Editor of the IEEE Communications Magazine and a member of the IEEE WCNC Steering Board.\\

AL\.{I} RIZA EKT\.{I} received his Ph.D. degree in Electrical Engineering from Department of Electrical Engineering and Computer Science at Texas A\&M University in 2015. He is with Grid Communications and Security Group of Electrification and Energy Infrastructure Division at Oak Ridge National Laboratory.\\

HALIM YANIKOMEROGLU [F] (halim@sce.carleton.ca) is a full professor in the Department of Systems and Computer Engineering at Carleton University, Ottawa, Canada. His research interests cover many aspects of 5G/5G+ wireless networks. His collaborative research with industry has resulted in 39 granted patents. He is a Fellow of the Engineering Institute of Canada and the Canadian Academy of Engineering, and he is a Distinguished Speaker for IEEE Communications Society and IEEE Vehicular Technology Society.
}
%%%%%%%%%%%%%%%%% BIOGRAPHIES %%%%%%%%%%%%%%%%%%%%%%

\balance
\end{document}